\documentclass[twocolumn]{article}

\usepackage{lineno,hyperref}
\usepackage{lipsum}
\usepackage{mathtools, cuted}
\usepackage{authblk}
\modulolinenumbers[5]
\usepackage[left=1.7cm,right=1.7cm,top=3.5cm,bottom=3.5cm]{geometry}
\usepackage{multicol}
\usepackage{xcolor}
\usepackage{cite}

\usepackage{amsmath,amssymb}
\usepackage[lang=en, warnundef]{jabbrv}
\DefineJournalAbbreviation{Review}{Rev}
\DefineJournalAbbreviation{Physica}{Phys}
\DefineJournalAbbreviation{Modern}{Mod}
\DefineJournalAbbreviation{Reports}{Rep}

		\title{Noise-induced nonreciprocal topological dissipative solitons in directionally coupled chains and lattices}
		
	\author[1,4]{D. Pinto-Ramos}
	\author[2]{K. Alfaro-Bittner}
	\author[3]{R. G. Rojas}
	\author[4]{M. G. Clerc}
	
\date{}

\affil[1]{\it Center for Advanced Systems Understanding (CASUS), Helmholtz-Zentrum Dresden-Rossendorf (HZDR), D-02826 Görlitz, Germany}	
\affil[2]{\it Universidad Rey Juan Carlos, Calle Tulip\'an s/n, 28933, M\'ostoles, Madrid, Spain}

\affil[3]{\it Instituto de F\'isica, Pontificia Universidad Cat\'olica de Valpara\'iso - Casilla 4059, Valpara\'iso, Chile}
\affil[4]{\it Departamento de F\'isica and Millennium Institute for Research in Optics, Facultad de Ciencias F\'isicas y Matem\'aticas,
	Universidad de Chile, Casilla 487-3, Santiago, Chile}

\begin{document}
\twocolumn[
  \begin{@twocolumnfalse}
    \maketitle	
\begin{abstract}
Nonreciprocal coupling can alter the transport properties of material media, producing striking phenomena such as unidirectional amplification of waves, boundary modes, or self-assembled pattern formation. 
{It is responsible for nonlinear convective instabilities in nonlinear systems that drive topological dissipative} solitons in a single direction, producing a lossless information transmission. 
Considering fluctuations, which are intrinsic to every macroscopic dynamical system, noise-sustained structures emerge permanently in time. 
Here, we study arrays of nonreciprocally coupled bistable systems exhibiting noise-sustained topological phase wall (or soliton) dynamics. 
The bifurcations between different steady states are analytically addressed, and the properties of the noise-sustained states are unveiled as a function of the reciprocal and nonreciprocal coupling parameters. 
Furthermore, we study critical points where the structures' characteristic size diverges with different power law exponents.
 Our numerical results agree with the theoretical findings.
\end{abstract}
 \end{@twocolumnfalse}
]
\subsection*{Introduction}
Natural systems in out-of-equilibrium conditions, subjected to both dissipation and injection of energy, are able to display a rich dynamical behavior. 
Examples are the synchronization phenomenon of coupled oscillators and chaotic systems \cite{pecora1990synchronization}, 
front propagation in population dynamics \cite{fisher1937wave,mollison1977spatial}, liquid crystal devices \cite{alvarez2019front, alfaro2020front}, 
or complex fluid flows \cite{ahlers1983vortex, fineberg1987vortex}, and the pattern formation, sometimes spatiotemporally complex, in chemical reactions \cite{lee1993pattern, maini1997spatial}, 
fluid dynamics \cite{cross1993pattern}, or active matter \cite{huber2018emergence, shankar2022topological}. 
Discrete nonlinear coupled systems compose a framework that has been successful for the description of dislocations in crystals \cite{aubry1983discrete}, arrays of coupled Josephson junctions \cite{braun2013frenkel}, 
excitable semiconductor lasers \cite{pammi2019photonic, alfaro2020pulse}, or coupled waveguide dynamics \cite{christodoulides1988discrete}, to mention a few. 
Sometimes, the discrete description even captures some details absent in continuous approximations of the dynamics \cite{clerc2011continuous}. 
Microscopic reversibility of time renders any coupled mechanical system to be reciprocal, a result known as the Maxwell-Betti reciprocity theorem \cite{maxwell1864calculation, coulais2017static, brandenbourger2019non}. 
However, the injection and dissipation of energy wisely applied could lead to a nonreciprocal response of coupled systems in the dynamic regime \cite{brandenbourger2019non,chen2021efficient,nassar2020nonreciprocity}, 
and not only statically \cite{coulais2017static}. The net effect can be the rupture of the space-reflection symmetry along an engineered direction \cite{brandenbourger2019non, nassar2020nonreciprocity}, 
causing the system to have interesting responses such as the uni-directional amplification of waves \cite{brandenbourger2019non}, 
boundary states \cite{shankar2022topological}, directed thermal flow \cite{xu2022diffusive}, self-assembled pattern formation \cite{pintoramos2021nrcisa}, 
giant boundary layers \cite{pinto2023giant}, or directed light propagation \cite{aguilera2024nonlinear}. 
Interestingly, complex natural systems such as vegetation in dry ecosystems have been modeled employing nonreciprocal interactions, 
which induces permanent stripe pattern dislocations \cite{pinto2023topological} and coexisting stripe oblique orientation domains \cite{hidalgo2024nonreciprocal}.
All these systems are subjected to noise and generally are nonlinear; thus, they are prone to the phenomenon of noise-sustained structures \cite{deissler1985noise}. 
This phenomenon, which creates and sustains nonlinear structures such as phase walls, defects, or patterns, occurs when the system has a spatial region where noise has enough energy to drive these structures \cite{aguilera2022thermal}. 
The mechanism is as follows: due to the boundary conditions, a boundary layer is established in the system, and perturbations grow easily inside it. 
Then, noise can create these structures in the boundary layer, where they are usually energetically cheaper to make than in the system bulk. 
Finally, for this dynamic to be persistent, advective flow (which can be effectively induced by nonreciprocal interactions) pushes the created structures into the bulk and then out of the system, 
at the same time, new ones are created at the boundary layer, forming a permanent dynamic of noise-sustained structures \cite{deissler1985noise}. 
This mechanism propagates complex nonlinear structures across the whole system, such as hydrodynamic vortices \cite{ahlers1983vortex, babcock1991noise}, 
turbulent flow \cite{deissler1985noise}, nematic liquid crystal patterns \cite{santagiustina1997noise, agez2006two}, and even topological defects \cite{clerc2015recurrent}. 
Noise cannot only sustain new dynamic regimes, but it can alter the existing ones in deterministic systems \cite{santagiustina1997noise}, 
and change the bifurcation curves of the underlying steady states \cite{agez2008universal}. 

In this work, we analyze the dynamics of nonlinear systems subjected to nonreciprocal coupling in chains and lattices. 
We also analyze how the bifurcation diagram of a damped nonreciprocal Frenkel-Kontorova prototype model is modified when noise is included. 
We test the robustness of the previously predicted self-assembled patterns and nonlinear waves of Ref. \cite{pintoramos2021nrcisa}. 
These structures are formed by topological dissipative solitons that carry information in a lossless way \cite{veenstra2024non}. 
The pattern of equispaced topological dissipative solitons of alternating signs remains stable for moderate nonreciprocity levels 
with a threshold that depends on the noise intensity. 
In addition, the absolute-convective instability exhibited in this discrete system gives rise to noise-sustained domains with a rich spatiotemporal dynamic.
\begin{figure}[]
	\centering
	\includegraphics[width=0.95\columnwidth]{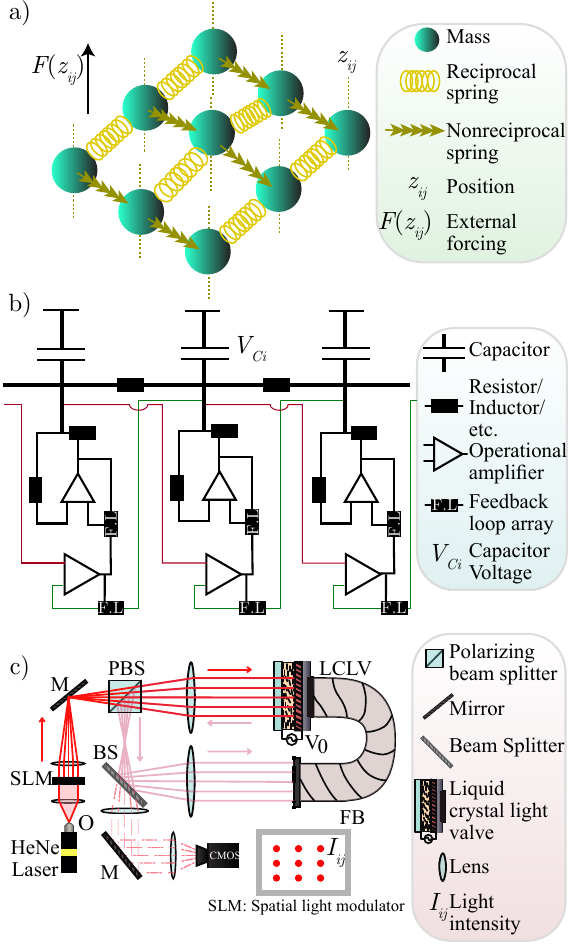}
	\caption{Discrete systems with nonreciprocal coupling. a) Membrane model of masses and their height $z_{ij}$, 
	one of the directions of the grid is coupled with a nonreciprocal material (indicated by arrows).  b) Electrical {circuits,
	the} voltage of each capacitor ($V_{Ci}$) can be nonreciprocally coupled employing operational amplifiers (OA). 
	The feedback loop array connected to each OA gives {the specific nonreciprocal coupling,} 
	and the nonlinear response is given by the circuit connected to the capacitor. c) Optical example.  
	A liquid crystal light valve (LCLV) reflects incoming light that is externally modulated by a spatial light modulator (SLM); 
	the reflection is translated and injected into the device through a {photo-conductive} plate. 
	The translation results in nonreciprocal coupling between the cells ($I_{ij}$) induced by the SLM, {for more details see Ref.~\cite{aguilera2024nonlinear}}.}
	\label{F1}
\end{figure}

\subsection*{Theoretical description}

Physical systems that could display discrete nonlinear dynamics of nonreciprocally coupled elements are not usual, 
but several experimental setups have recently been developed due to their striking properties. 
Mechanical systems have been constructed employing robotic metamaterial chains \cite{brandenbourger2019non,veenstra2024non}. 
Similarly, a chain of nonlinear damped oscillators has been studied theoretically under the aforementioned nonreciprocal coupling using the {Frenkel-Kontorova model~\cite{braun2013frenkel}. 
The reciprocal Frenkel-Kontorova model was initially proposed to describe a chain of classical particles with nearest-neighbor interactions, subject to a periodic on-site substrate potential.
Later, it described various physical phenomena, including dislocations, dynamics of adsorbed layers on surfaces, domain walls in magnetically ordered structures, and even DNA chains.
This chain} can be extended laterally with reciprocal or non-reciprocal couplers to form lattices, as illustrated in Fig.~\ref{F1} a).  
Apart from mechanical systems, nonreciprocal couplers can be constructed for the dynamics of current or voltage in electrical circuits. 
There, nonlinear circuits showing bistability or chaos could be connected through operational amplifiers in chains or 
other arrays to form both unidirectional or non-reciprocal coupled circuits \cite{tamavsevivciute2008analogue,matias1998transient, zheng2001collective}. 
An example is illustrated in Fig.~\ref{F1} b). On the other hand, nonreciprocity in nonlinear optical devices could 
be induced by translated optical feedback \cite{del2012effects, alvarez2020transition,aguilera2024nonlinear}, as shown in Fig.~\ref{F1} c).

We start by assuming a discrete chain of particles subjected to a periodic (or bistable) potential and nearest neighbor linear elastic coupling, the Frenkel-Kontorova model \cite{braun2013frenkel}. Moreover, we allow the coupling to be nonreciprocal.
{A Lagrangian $\cal{L}$ can describe the system, reading }
\begin{equation}
	{\cal L} = \sum_i \left[ \frac{\dot{\theta}_i^2}{2} - \omega^2 \cos \theta_i
	-  \frac{D-\alpha}{2}(\theta_{i+1}-\theta_i)^2  \right]e^{\mu t}\Lambda^{i},
	\nonumber
	\label{Eq-Lagrangian}
\end{equation}
where $\Lambda \equiv (D-\alpha)/(D+\alpha)$. $\omega^2$ characterizes the in-site potential. 
$D$ corresponds to the reciprocal elastic constant and $\alpha$ the nonreciprocal part of it. 
$\mu$ is the viscous damping coefficient modeling dissipation. 
Following the least-action principle and imposing the overdamped regime~\cite{pintoramos2021nrcisa}, 
the equation of motion for a nonreciprocally coupled chain is 
\begin{equation}
	\dot \theta_i= \omega^2 \sin\theta_i + (D-\alpha)(\theta_{i+1}-\theta_i)-(D+\alpha) (\theta_i-\theta_{i-1}).
	\label{Eq-CoupledChain}
\end{equation}
Note that $\theta$ is measured such that the equilibria $\theta=0$ and $\theta=\pm \pi $ are unstable and stable, respectively. 
The term $\omega^2 \sin \theta_i$ could be replaced with any nonlinear on-site force exhibiting bistability. 
Dynamics of equation (\ref{Eq-CoupledChain}) have been described when imposing a fixed boundary condition $\theta_0 =0$ \cite{pintoramos2021nrcisa}. 
This restriction favors the formation {of nonlinear waves between stable and unstable states, \textit{FKPP} fronts~\cite{van2003front}, }
with different velocities depending on their propagation direction. 
Nonreciprocal coupling could induce the system (\ref{Eq-CoupledChain}) to have a kink and anti-kink (topological  {dissipative} solitons of opposite charge) emission 
either from a former \textit{FKPP} front or from the upstream boundary (forming a self-assembled pattern) 
depending on the nonreciprocity level $\alpha$ deterministically \cite{pintoramos2021nrcisa}. 

We extend the chain transversally while still analyzing a scalar field, forming a lattice. 
Additionally, we include the effect of fluctuations in a highly dissipative environment. The nonreciprocally coupled lattice obeys
\begin{eqnarray}
	\dot \theta_{ij}&=& \omega^2 \sin\theta_{ij} + (D-\alpha)(\theta_{i+1j}-\theta_{ij})\nonumber\\
	&&-(D+\alpha) (\theta_{ij}-\theta_{i-1j})+D_{\perp}(\theta_{ij+1}-\theta_{ij})\nonumber\\ 
	&&-D_{\perp}(\theta_{ij}-\theta_{ij-1})+ \sqrt{\Gamma}\xi_{ij}(t),
	\label{Eq-CoupledChain_2D}
\end{eqnarray}
{where $\theta_{ij}$ accounts for the dynamics of the $ij$-element of the lattice,} 
$D_{\perp}$ represents a transversal (to the direction with nonreciprocal coupling) {reciprocal} coupling in the {bistable} chain, 
for simplicity, we consider it constant ($D_\perp = D$), 
but it could be a nonlinear function of $\theta_{ij}$, {as well,} depending on the underlying physics. 
$\Gamma$ measures the intensity {level} of the randomly generated fluctuations $\xi_{ij}(t)$ (in time and in sites), which is characterized by white noise statistics,
{that is, the fluctuations are characterized by a Gaussian stochastic process with a zero mean value and without correlation. 
The dynamics of Eq.~(\ref{Eq-CoupledChain_2D}) are similar to Eq.~(\ref{Eq-CoupledChain}) and will be dominated by moving phase walls. }

{To study noise-sustained structures in the system,} we implement a boundary condition that favors the formation of boundary layers. 
Then, we impose $\theta_{0j} = 0$ {for all $j$} (the unstable equilibrium). 
The other end of the lattice has free boundary conditions such that $\theta_{(N+1) j} = \theta_{Nj}$ {for all $j$} (where $N$ is the {length} of the system in the $i$ direction), 
but the results hold for other choices. 
In the transversal direction to nonreciprocal coupling, we impose periodic boundary conditions $\theta_{i(L+1)} = \theta_{i0}$ {for all $i$}
(where $L$ is the length of the system in the $j$ direction).

\subsection*{Results}
\subsubsection*{Numerical observations}
\begin{figure*}[ht!]
	\centering
	\includegraphics[width=2\columnwidth]{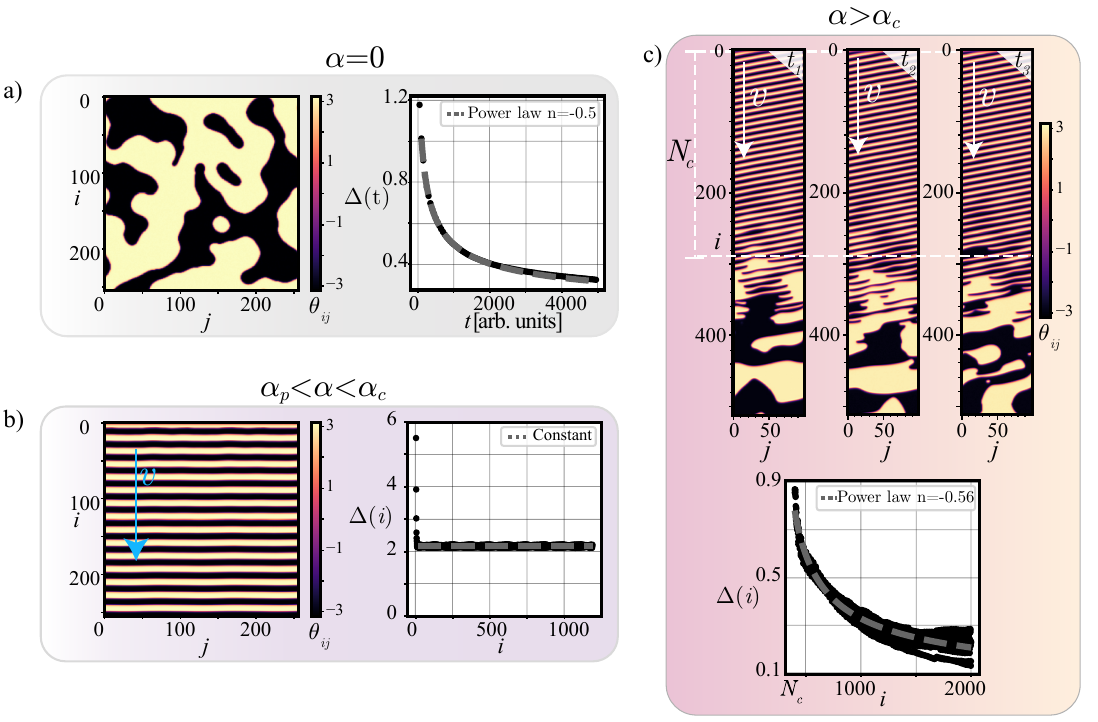}
	\caption{Dynamics of the nonreciprocally coupled Frenkel-Kontorova lattice. a) Case $\alpha=0$. 
	The left panel shows a snapshot of the dynamics in the absence of nonreciprocity; 
	the right panel depicts how the phase wall total perimeter ($\propto \Delta(t)$) obeys a coarsening law in time ($\Delta(t) \propto t^{n}$) with exponent $n=-0.5$. 
	b) Case $\alpha_p<\alpha<\alpha_c$, where self-assembled patterns can be observed despite the existence of fluctuations. 
	A snapshot is shown in the left panel, and the behavior of the {macroscopic} parameter $\Delta(i)$ is given in the right panel. 
	The light blue arrow indicates the direction of {propagation} of the pattern, with velocity $v$. c) Case $\alpha>\alpha_c$. 
	The top panel shows snapshots of the dynamic ($t_1<t_2<t_3$), where a boundary layer of size $N_c$ exhibiting the patterned state is followed by intricate permanent advected phase walls. 
	The bottom panel shows the coarsening law of the $\Delta(i)$ variable, which shows coarsening dynamics ($\Delta(i)\propto i^{n}$) with exponent $n=-0.5$. 
	The white arrows indicate the {propagation} direction of the pattern and the phase walls with velocity $v$. All simulations where performed with parameters $\omega=1$, $D=0.3$, and $\sqrt{\Gamma}=0.1$.}
	\label{F2}
\end{figure*}

To analyze the dynamics of Eq.~(\ref{Eq-CoupledChain_2D}), it is convenient to define a {macrosopic} parameter quantifying the presence of interfaces connecting the $\theta=\pm \pi$ equilibria, 
often called phase walls, domain walls, or kinks, which can not be destroyed by smooth transformations of the variable; 
thus, they are topological, sometimes called topological {dissipative} solitons (due to the soliton-like profile of their spatial derivative). 
A possible choice corresponds to the parameter $\Delta(t) = \pi^2 - \frac{1}{N}\sum_{ij} \theta_{ij}^2(t)$. 
A homogeneous solution with $\theta_{ij}=\pi$ or $\theta_{ij}=-\pi$ gives a value $\Delta =0$. 
An interface necessary crosses the zero, so its presence increases the $\Delta$ value. 
It is not hard to convince oneself that for a 1-dimensional array, $\Delta \propto (\text{total number of interfaces})$, 
and that in a 2-dimensional lattice $\Delta \propto (\text{total interface length})$, given that the interfaces are identical to each other, 
which is granted in our system due to homogeneity of space (the equation is independent of $(i,j)$ explicitly). 
In the presence of nonreciprocal interactions in the $i$ direction, in steady state operation, it is convenient to define $\Delta$ 
as a function of $i$ instead of time, such that $\Delta(i)=  (1/T) \int_0^T \left[\pi^2 - \sum_{j} \theta_{ij}^2(t)\right] dt$.

Numerical simulations of model Eq.~(\ref{Eq-CoupledChain_2D}) display the behaviors summarized in Fig.~\ref{F2}. 
In the absence of nonreciprocity, $\alpha=0$, domain wall dynamics is observed, where a coarsening law for the parameter $\Delta(t)$ is found, 
with a characteristic exponent of $-1/2$ ($\Delta\sim t^{-1/2}$). This exponent is explained by phase ordering kinetics, see the appendix for details.
If $D \lesssim \omega^2$ and $\alpha_c>\alpha>\alpha_p$, 
we are found in a region where a self-assembled pattern of phase walls (or kinks) with alternating topological charge emerges; 
$\alpha_p$ corresponds to the onset of the pattern state, and $\alpha_c$ {is} the onset for noise sustained structures. 
In this region, either measuring $\Delta(t)$ or $\Delta(i)$ gives a constant value due to the periodic behavior of the system. 
Finally, one has the $\alpha>\alpha_c$ region, where the above-mentioned pattern (or the homogeneous state if $D\gtrsim\omega^2$) 
loses its stability and gives rise to a permanent domain wall dynamical {behaviors;} interestingly, a coarsening-like law is observed for $\Delta(i)$, 
with the same exponent $-1/2$. One can readily see that in the $\alpha>\alpha_c$ case, a boundary layer is formed, 
which size we call $N_c$. 
The coarsening law for $\Delta(i)$ is valid only for $ i>N_c$, because for $ i <N_c$ 
the system is either on a homogeneous state ($D\gtrsim \omega^2$) or in the pattern state $(D\lesssim \omega^2$). 
The boundary layer size depends on the coupling parameters $\alpha$ and $D$, and also on the noise intensity level $\Gamma$ 
which is responsible for destabilizing the pattern (or homogeneous state). 

 \begin{figure}[]
	\centering
	\includegraphics[width=0.9\columnwidth]{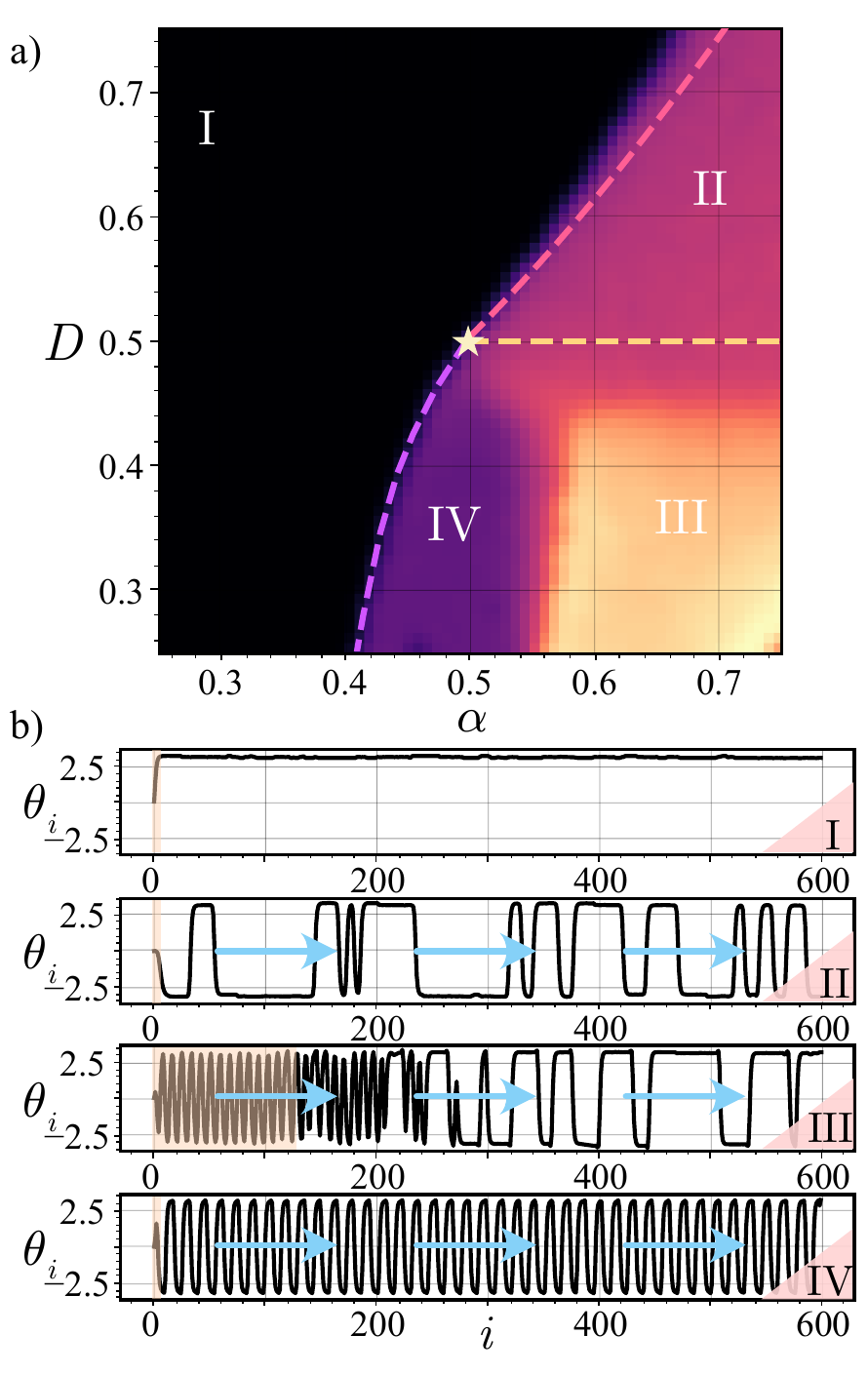}
	\caption{Phase diagram of Eq.~(\ref{Eq-CoupledChain_2D}) for planar solutions and their typical behavior depending on the coupling parameters $D$ and $\alpha$. 
	a) {The phase diagram of Eq.~(\ref{Eq-CoupledChain_2D}) obtained numerically.}
	Four regions could be differentiated by analyzing the domain walls' number fluctuations and the boundary layer size. 
	The dashed lines correspond to transition curves obtained in the absence of noise. 
	b) {Pannels illustrate the typical} behavior of each region. 
	Region~I corresponds to the homogeneous, stable state; one observes no domain walls and a thin boundary layer (the shadowed region {on the left flank}). 
	Region~II shows permanent domain wall dynamics (whose number fluctuates) with {a thin boundary layer}. 
	In {Region}~III, the boundary layer enlarges drastically up to $N_c$, where the permanent domain wall dynamics (whose number fluctuates) emerge. 
	{Finally, in Region}~IV, stable self-assembled patterns can be observed; the boundary layer is thin, and the number of phase walls is constant. 
	All simulations have parameters $\omega=1$ and $\sqrt{\Gamma}=0.1$.}
	\label{F3}
\end{figure}

Despite the complexity arising from  {complex-shaped walls, one numerically obtains that their} total perimeter obeys a simple law for its evolution in time. 
For enough nonreciprocity, this behavior is translated to space. 
Intuitively, this can be explained by looking at the continuum limit of the model, 
where the nonreciprocal coupling term transforms into a linear advection term; 
thus, phase walls drift with constant velocity. {At the same time,} they relax according to their interaction laws.
 
In the case $\alpha \neq 0$, {the equation exhibits a clear preference for} the $i$ direction, and modulations in $j$ are perturbations. 
For simplicity, as the main dynamics are longitudinal, 
we use equation~(\ref{Eq-CoupledChain_2D}) in the case $\theta_{j\pm1}=\theta_j$ (planar solutions) for computational efficiency and focus on domain wall dynamics, 
as they mediate the behavior of the system. 
The phase walls' dynamical regimes are depicted in Fig.~\ref{F3} as a function of the coupling parameters, where we observe rich dynamics converging to homogeneous states, 
patterns, or permanent phase wall dynamics with different boundary layers. 
We can recognize four different behaviors for the system dynamics, corresponding to Regions I, II, III, and IV of the phase diagram in Fig. \ref{F3}. 
Region I corresponds to the system achieving a homogeneous state $\theta_i= \pi$ or $\theta_i= -\pi$, due to the boundary conditions, a small boundary layer is observed. 
In {Region} II, it is observed that from a small boundary layer, phase walls are permanently created, and one observes a steady state with a nonuniform distribution of them. 
In {Region} III, the boundary layer extends and takes a big portion of space; 
the boundary layer is similar to a pattern that becomes unstable at the mean distance $N_c$, and for $i>N_c$, 
the permanent domain wall dynamic is recovered. 
Finally, in {Region} IV, one can observe the self-assembled pattern as the steady state.

To understand how these different behaviors are distributed in the parameter space analytically, 
an analysis of the \textit{FKPP} front dynamics is performed, and the boundaries of the phase diagram are obtained approximately.  
A {valuable} tool for this corresponds to the calculated mean velocity of \textit{FKPP} fronts, $v(\alpha, D)$, presented in \cite{alfaro2017pi,pintoramos2021nrcisa}. 
In the deterministic case $\Gamma=0$, it is found that depending on the parameters, the {the shape of \textit{FKPP} front as a function the space} 
can be monotonous or nonmonotonous, 
and each of its forms can suffer an absolute convective instability. 
Nonmonotonous fronts arise due to a modulational instability of the monotonous front solution. 
In the case {of noisy system}, $\Gamma \neq 0$, one would assume that all \textit{FKPP} fronts are destroyed due to their fragility near the unstable equilibrium $\theta =0$, and thus become irrelevant; 
however, their properties leave an imprint {on} the system behavior, as seen in the next section.
 \subsubsection*{Analytical predictions}

\begin{figure}[ht!]
 \includegraphics[width=0.95\columnwidth]{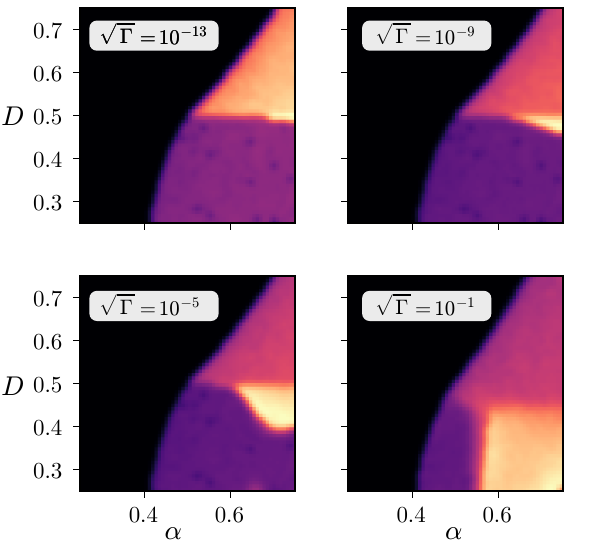}
 	\caption{Effect of noise in the dynamical regions of the phase diagram depicted in Fig. \ref{F3} for planar solutions of Eq. \eqref{Eq-CoupledChain_2D}. 
	One can see that for vanishing noise, one approaches the curves predicted by the deterministic model. 
	For increasing noise, the regime of self-ensembled phase walls has a smaller domain in parameter space, while the one of incoherent emission of the topological defects enlarges. }
 	\label{F4}
 \end{figure}

 Planar \textit{FKPP} front solutions of Eq.~(\ref{Eq-CoupledChain_2D}) in the deterministic limit $\Gamma=0$ can be analyzed with the ansatz {for the front tail}
 $\theta_i= \epsilon \exp\left[\sqrt{-1}(ki-\Omega t) \right]$, where $k$ and $\Omega$ are the wavenumber and angular frequency (allowed to be complex \cite{van2003front}), obtaining the linear growth relation
 \begin{equation}
 	-\sqrt{-1} \Omega(k) = 1-2D + 2D\cos(k)- 2\alpha\sqrt{-1}\sin(k).
 \end{equation}
Then, we can apply the self-consistency equations for the existence of \textit{FKPP} fronts propagating with velocity $v$ and wavenumber $k^c$ \cite{van2003front}, 
reading $d\Omega/dk |_{k^c}=0$ and $v= \text{Im} \Omega / \text{Im}k^c$. When using the boundary condition $\theta_0=0$, 
{analyzing solutions with $\text{Im}k^c<0$ is especially interesting, as these fronts could undergo an absolute-convective instability \cite{pitaevskii2012physical,chomaz1992absolute}.}
This instability occurs at the condition $v(\alpha, D, k^c)=0$, which gives the light purple (light) and yellow (lightest) dashed curves in the phase diagram of Fig- \ref{F3}. 
The dashed light purple (light) curve corresponds to the absolute-convective threshold for monotonous ($\text{Re}k^c=0$) \textit{FKPP} fronts and has the analytical formula $D=\omega^2/4 + \alpha^2/\omega^2$.
 On the other hand, the dashed yellow (lightest) curve is the absolute-convective threshold for non-monotonous ($\text{Re}k^c\neq0$) \textit{FKPP} fronts and corresponds to the curve $D=\omega^2/2$. 
 One can clearly observe that fluctuations affect the transition curves; this is due to the fragility of the \textit{FKPP} front that forms close to the boundary $\theta_0=0$. 
 As a consequence of fluctuations, Region II supports noise-sustained structures \cite{deissler1985noise}, from the upstream boundary, $\theta_0=0$ 
 perturbations grow and are advected {due to} the state $\theta=0$ being convectively unstable. 
 A similar behavior is observed in Region III, where a spatially periodic boundary layer gives rise to the noise-sustained structures. 
 Fluctuations modify the landscape in the $(D,\alpha)$ parameter space compared to the deterministic case seen in Fig. 2 of Ref. \cite{pintoramos2021nrcisa}. 
 In this reference, for $\Gamma=0$, Region IV extends over Region III, Region II shows $\theta=0$ as the homogeneous steady state, and the transition curves match exactly with the predictions given. 
 We reproduce this in Fig. \ref{F4} considering a small value of {noise level intensity} $\sqrt{\Gamma}=10^{-13}$.

{The extent to which the noise term affects the phase diagram depends on its intensity level $\Gamma$.}
A less intense noise will be less capable of creating phase walls in Region II, so the average distance between them increases. 
Likewise, a less intense noise has a weaker effect on the pattern of self-ensembled phase walls. 
Indeed, Region III would be reduced as the boundary layer size $N_c$ would increase and eventually reach the system size. 
This enlargement of the boundary layer size emerges because the perturbation does not have enough time to disarm the pattern, 
in particular, boundary layer size behaves as $N_c \sim -\log \Gamma$, similar to the case of giant boundary layers in Ref. \cite{pinto2023giant}. 
The effect of the noise on the transition curves is depicted in Fig.~\ref{F4}, where one can observe that the convective instability curve and the transition to self-ensemble of phase walls are robust. 
However, the domain of the self-ensemble shrinks as the region for incoherent emission of defects enlarges. 
Nevertheless, one can still observe the controlled emission of the phase walls for high noise values; this is due to their nonlinear origin as a limit cycle solution.

We note that the curve separating Regions I and IV, the dark purple (dark) dashed curve in Fig. \ref{F3}, is not predicted with the theory of \textit{FKPP} front propagation. 
Here, we report its origin. To understand this curve in a tractable manner, let us consider the limit of a few pendulums, with two of them being enough (see the Appendix for details). 
The equations for the two pendulums $(\theta_1, \theta_2)$ will have multiple equilibria representing {their rest position.} 
One trivial solution is $(\theta_1, \theta_2)=(0, 0)$, which is always unstable; 
generally, four other solutions exist: two of them represent symmetric stable configurations $(\theta_1, \theta_2)\neq(0,0)$, and the other two are unstable ones. 
The purple dashed curve of the phase diagram in Fig. \ref{F3} corresponds to saddle-node on invariant curve (SNIC) bifurcation \cite{izhikevich2007dynamical, strogatz2018nonlinear}; 
the four equilibria mentioned above annihilate pairwise in a saddle-node bifurcation at $\alpha=\alpha_p(D)$, 
however, the heteroclinic orbits close in a loop, forming a stable limit cycle for $\alpha>\alpha_p(D)$. 
The curve is approximated implicitly by the equations (see the Appendix for details)
\begin{eqnarray}
	\theta = \frac{2D \left[\theta -\frac{\sin \theta}{D+\alpha_p}\right]-\sin\left(\theta- \frac{\sin \theta}{D+\alpha_p}\right)}{D-\alpha_p}, \nonumber \\
	1=\frac{1}{D-\alpha_p}\left(1-\frac{\cos \theta}{D+\alpha_p}\right)\left(2D-\cos( \theta- \frac{\sin \theta}{D+\alpha_p})\right). \label{pattern_curve}
\end{eqnarray}
Surprisingly, the curve predicted with a two-pendulum theory fits perfectly the numerical results. 
Note that as this bifurcation gives birth to a limit cycle, it explains the robustness of the pattern of self-assembled structures against fluctuations.

 Finally, we analyze the scaling law for the total perimeter of domain walls in the system. 
 Clearly, the dynamic in the bulk is dominated by phase wall interactions. It is known that in Eq.~\eqref{Eq-CoupledChain} bistable fronts (or phase walls) are weakly interactive, 
 as a consequence of both states $\theta =\pm \pi$ being favorable to the system. 
 The interaction can be quantified if one employs an approximation in the continuum limit \cite{alfaro2017pi, alfaro2019traveling}, resulting in that a slightly curved front relaxes diffusively to a straight line \cite{pismen2006patterns}. 
 Considering this, one obtains the characteristic exponent $n=-0.5$ for the relaxation of the perimeter of phase walls (see the Appendix for details). 
This exponent characterizes the noise-sustained structures of Regions II and III of the phase diagram. 
One path to reach these dynamic states is to increase the nonreciprocal coupling parameter $\alpha$, 
when doing so, depending on the value of $D$, we will observe or not the self-assembled patterns. 
 
\begin{figure}[ht!]
 	\centering
 	\includegraphics[width=0.8\columnwidth]{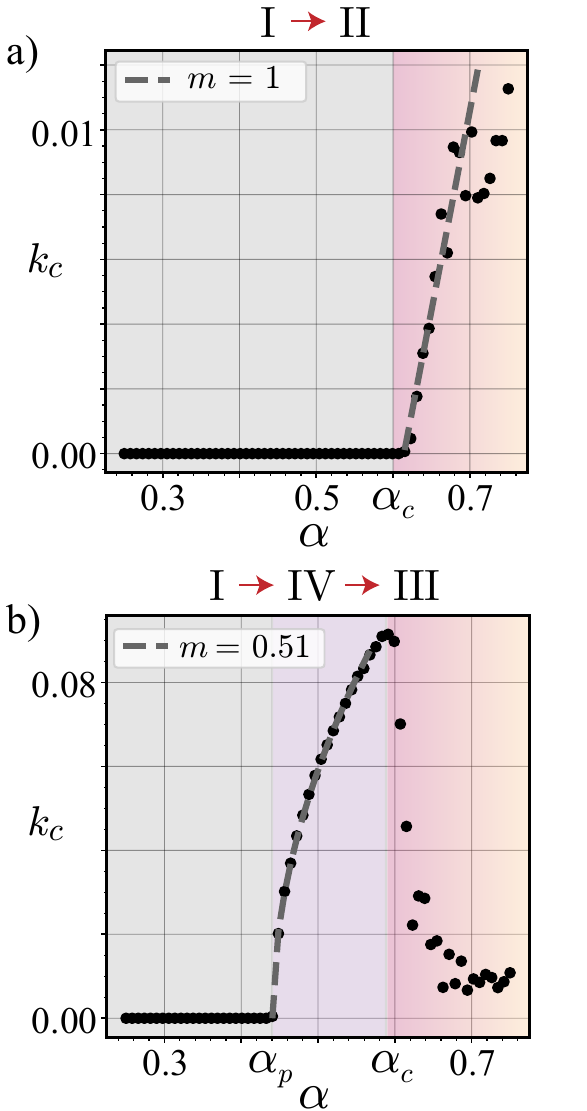}
 	\caption{Behavior of the characteristic length of the system structures after the bifurcations. a) Transition from Region I to II. For $\alpha<\alpha_c$ the system is homogeneous, thus the dominant Fourier mode is zero. For $\alpha>\alpha_c$, the dominant mode increases linearly with $\alpha$, that is, $k_c \propto (\alpha-\alpha_c)^m$, with $m=1$. b) Transitions from Region I to IV, and IV to III. For $\alpha<\alpha_p$ the system is homogeneous. For $\alpha_c>\alpha>\alpha_p$, one observes the pattern state, which exhibits a characteristic mode behaving as $k_c \propto (\alpha-\alpha_p)^m$, with $m=0.5$. For $\alpha>\alpha_c$, one enters Region III, where the characteristic length of the boundary layer, the pattern inside it, and the randomly distanced phase walls coexist.}
 	\label{F5}
 \end{figure}

 A useful parameter to monitor is the characteristic size of the phase wall pairs that form both the self-assembled pattern or the noise-sustained structure \cite{pintoramos2021nrcisa}, 
represented by the characteristic wavenumber of the Fourier power spectrum. 
We compute it numerically as a function of the nonreciprocal coupling $\alpha$, shown in Fig. \ref{F5}. When increasing $\alpha$ for $D>D_c$, 
we find that the characteristic size of structures in the system behaves as $\lambda\sim 1/(\alpha-\alpha_c)$, with $\alpha_c= \sqrt{D\omega^2-\omega^4/4}$. 
This scaling is explained when computing the average number of phase walls emitted from the thin boundary layer in {Region} II, 
which scales as $n\sim \alpha-\alpha_c$ (see Appendix for details), 
using that $\lambda n\sim N$ (the number of elements) we recover the scaling for the characteristic size behavior. 
On the other hand, if $D<D_c$ when increasing $\alpha$, one first encounters the region of self-assembled patterns, 
for which the wavelength scales as $\lambda\sim 1/(\alpha-\alpha_c)^{1/2}$ with $\alpha_c$ obtained from Eq.~\eqref{pattern_curve}. 
This is explained by the nature of the SNIC bifurcation \cite{izhikevich2007dynamical, strogatz2018nonlinear}, 
for which the period of oscillations scales as $T\sim 1/(\mu-\mu_c)^{1/2}$ with $\mu$ the bifurcation parameter. 
 
\subsection*{Conclusion}
The phase diagram for a Frenkel-Kontorova chain or lattice is unveiled when subjected to {simultaneously} nonreciprocal coupling and fluctuations. 
Noise-sustained structures are observed {due to} noise and nonreciprocity, which produce an effect similar to advection in {continuum} systems. 
These noise-sustained structures exhibit coarsening dynamics in the co-mobile reference frame of the respective created structures. 
Thus, the temporal coarsening is transformed into a {\it spatial coarsening}. 
These complex dynamical behaviors are mediated by the nonreciprocity parameter and the reciprocal coupling parameter $\alpha$ and $D$, respectively. 
Furthermore, they are characterized by the average size of structures versus parameters.

In the case of small fluctuations, one can consider the reciprocal coupling parameter $D_c\approx \omega^2/2$ a special one, 
as it separates two different paths in parameter space that lead to the regime of randomly emitted domain walls. For $D>D_c$, 
when increasing the nonreciprocity parameter $\alpha$, we observe a sharp transition from the homogeneous state to the noise-sustained state of phase walls (also called topological solitons). 
On the other hand, for $D<D_c$, we have {particular} values of $\alpha_{p}<\alpha<\alpha_{c}$ for which a self-assembled pattern stabilizes, 
although it becomes unstable for enough nonreciprocity. 
The pattern is resilient to noise and has a well-defined wavelength given by the system parameters. 
In this regime, a perfect array of phase walls, or topological solitons, is emitted from one boundary and transmitted to the other in a self-sustained fashion without deformations. 

To conclude, this work explored the effects of noise over nonreciprocally coupled chains of nonlinear systems. 
Analytical results regarding the bifurcations of a prototypical model are obtained and contrasted numerically. 
Noise considerably affects the bifurcations mediated by \textit{FKPP} fronts, as these nonlinear waves are fragile against additive fluctuations. 
On the other hand, we found that an infinite period bifurcation gives rise to a spatiotemporal-periodic state that could not be explained before, 
we unveiled that its topological nature (in phase space) protects it against fluctuations, making it robust. 

The different states supporting continuous emission of phase walls (the noise sustained Regions II and III, 
and the pattern Region IV) are characterized by a coarsening dynamic for the total perimeter of walls. 
Similarly, the average size of the structures against the nonreciprocity parameter is unveiled as a function of the nonreciprocity while crossing the bifurcation curves. 
Our results can enlighten the path to nonlinear nonreciprocal device characterization and operation in fluctuating environments.

\subsection*{Acknowledgments}
D.P.-R. acknowledges the financial support of ANID National Ph.D. scholarship 2020-21201484. 
M.G.C. acknowledges the financial support of ANID-Millennium Science Initiative Program-ICN17$\_$012 (MIRO) and FONDECYT project 1210353. 
This work was partially funded by the Center of Advanced Systems Understanding (CASUS), 
which is financed by Germany’s Federal Ministry of Education and Research (BMBF) 
and by the Saxon Ministry for Science, Culture, and Tourism (SMWK) with tax funds on the basis of the budget approved by the Saxon State Parliament.

\subsection*{Appendix}

\subsubsection*{A. Saddle node on invariant curve bifurcation}

{The} origin of the bifurcation separating Regions I and IV is in the upstream boundary, which we realized by numerical inspection. 
We observe that only a few particles acquire self-sustained dynamics, and the rest of the chain accommodates its perturbations. 
Motivated by this, we consider the limit of two-particle dynamics described by the equations
\begin{eqnarray}
	\dot{\theta}_1= \sin \theta_1 -2D\theta_1 +(D-\alpha)\theta_2,\nonumber\\
	\dot{\theta}_2=\sin \theta_2 -(D+\alpha)\theta_2 +(D+\alpha)\theta_1.
\end{eqnarray}

This simple dynamical system supports various equilibrium states. In particular, $(\theta_1, \theta_2)=(0,0)$ corresponds to an unstable node. 
The solutions representing the thin boundary layer in Region I correspond to $(\theta_1, \theta_2) = \pm(\phi_1, \phi_2)$; 
however, the solution for these angles is not unique. Indeed, hyperbolic points representing unstable boundary layers exist. 
When these equilibria cease to exist by saddle-node bifurcation, a limit cycle of an infinite period is born from the heteroclinic orbits of the degenerated equilibrium points at the bifurcation, 
namely, a saddle-node on an invariant curve (SNIC) bifurcation \cite{izhikevich2007dynamical}.

To determine the bifurcation point, we need to ask the parameter point for which $\phi_{1,2}$ ceases to have solutions, 
that is, they become complex-valued. Solving for $\phi_2=\phi$ leads to

\begin{equation}
\phi = \frac{2D \left[\phi -\frac{\sin \phi}{D+\alpha}\right]-\sin\left(\phi- \frac{\sin \phi}{D+\alpha}\right)}{D-\alpha}.\label{phi}
\end{equation}

\begin{figure}[ht!]
 	\centering
 	\includegraphics[width=\columnwidth]{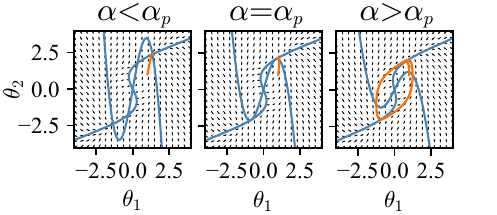}
 	\caption{System of two pendula for values of the nonreciprocity before, at, and after the SNIC bifurcation. 
	Blue lines correspond to the nullclines of the system of equations. 
	Orange lines correspond to the trajectories starting from the initial condition $(\theta_1, \theta_2)=(1,1)$. }
 	\label{F6}
 \end{figure}

Close to the saddle-node bifurcation, two intersections of the above curves exist. Just at the saddle-node; 
they become tangent. This is the condition for the SNIC bifurcation. Thus, one solves Eq.~\eqref{phi}, obtaining $\phi^*(D,\alpha)$; 
then, deriving Eq.~\eqref{phi} to impose the saddle-node bifurcation, {one gets}
\begin{equation}
1=\frac{1}{D-\alpha_p}\left(1-\frac{\cos \phi^*}{D+\alpha_p}\right)\left(2D-\cos( \phi^*- \frac{\sin \phi^*}{D+\alpha_p})\right). \label{alpha_p}
\end{equation}
Equation \eqref{alpha_p} determines the bifurcation curve in the phase diagram $(D, \alpha)$ that leads to the permanent spatiotemporal-periodic emission of topological solitons. 
Then, the curve separating Regions I and IV is given by $D= D(\alpha_p)$, which is implicit in Eq. \eqref{alpha_p}. 
An example of the bifurcation is depicted in Fig. \ref{F6}.
\subsubsection*{B. Coarsening scaling exponent}

Scaling of the perimeter of the phase walls is an already-answered question regarding continuous systems in space. 
Considering that the transition to a continuum limit for equation (\ref{Eq-CoupledChain_2D}) is a matter of scaling of the parameters, 
the same results as in Ref.\cite{bray1994growth} can be applied. Our equation is for a single variable in two dimensions, 
then, the scaling rule for the characteristic length of the domains is $L(t)\sim t^{1/2}$. 
Naturally, in two dimensions, the number of these defects goes as $n_{\text{phase walls}}\sim 1/L^2$, 
thus, the total perimeter of phase walls escales as $P\sim N\cdot L = 1/L$ \cite{bray1994theory,bray1994growth}. 
This demonstrates the exponent observed in the $\alpha=0$ case.

For the $\alpha\neq 0$ case, the exponent observed this time in space can be attributed to the linear relationship between the drift velocity 
of the topological solitons and the nonreciprocity parameter $v\sim \alpha$ \cite{aguilera2024nonlinear}. 
Then, in the coarsening region (away from the boundary layer), the coarsening dynamic occurs in a mobile reference frame with speed $v$. 
Finally, one can interchange the role of time and space due to the linear relation between the time traveling and the distance crossed by a defect $t_{\text{travel}} =  x_{\text{crossed}}/v$, 
similar to the argument employed in Ref. \cite{pinto2023topological}. 
This explains that the same exponent holds, however, this time against space, for the perimeter of phase walls in the case $\alpha\neq 0$. 

\subsubsection*{C. Defects created in the boundary layer}

To derive an approximation for $n$ defects created at the boundary layer, we employ simple arguments. 
It is already known that at first order, nonreciprocity behaves like a linear advection \cite{aguilera2024nonlinear}. 
Then, the velocity of $FKPP$ fronts close to the convective instability is $v_{FKPP}\sim \alpha-\alpha_c$. 
Then, one can argue that the boundary layer size is established as the velocity of the $FKPP$ front and a characteristic timescale arising from the growth rate of perturbations given by noise. 
This timescale is just $\tau \sim -\log \Gamma$ (similar to the argument used in \cite{aguilera2022thermal}), {then $N_c \sim -(\alpha- \alpha_c)\log \Gamma$.}
The maximum number of zeros that can fit in the boundary layer is given by the characteristic wavenumber of the modes excited by noise times the boundary layer length. 
The linear growth relationship gives the characteristic wavenumber, obtaining approximately that $k_c\sim 1/\sqrt{D}$. 
Finally, using that $n =N_c k_c$ one obtains that $n_{\text{created}} \sim -(\alpha-\alpha_c)\log \Gamma/\sqrt{D}$. 
This serves as the initial condition for the annihilation that takes place at the polynomial rate $n\sim t^{-1}$ (in the two-dimensional case, for others, see \cite{bray1994theory,bray1994growth}). 
Moreover, the total number of phase walls in the system is directly proportional to the ones created at the boundary layer, 
explaining the obtained result for the characteristic domain size as a function of $\alpha$.



\end{document}